\documentclass[12pt]{article}

\usepackage{amssymb}
\usepackage{amsmath,amsthm, mathtools,commath}
\usepackage{graphics, color, comment}

\usepackage{graphicx}
\usepackage{epsfig}
\usepackage{makeidx}
\usepackage{kotex}
\usepackage{fullpage}
\usepackage{bm}

\usepackage[round]{natbib}

\newcommand{\biblist}{\begin{list}{}
{\listparindent 0.0cm \leftmargin 0.50cm \itemindent -0.50 cm
\labelwidth 0 cm \labelsep 0.50 cm
\usecounter{list}}\clubpenalty4000\widowpenalty4000}
\newcommand{\ebiblist}{\end{list}}

\usepackage{latexsym}
\usepackage{amsmath, amssymb, amsfonts, amsthm, bbm}
\usepackage{graphicx}
\usepackage{mathrsfs}

\newcommand{\bx}{\mathbf{x}}

\newcommand{\bkappa}{{\mbox{\boldmath$\kappa$}}}
\newcommand{\balpha}{{\mbox{\boldmath$\alpha$}}}
\newcommand{\bbeta}{{\mbox{\boldmath$\beta$}}}
\newcommand{\bphi}{{\mbox{\boldmath$\phi$}}}

\newcommand{\bgamma}{{\mbox{\boldmath$\gamma$}}}
\newcommand{\blambda}{{\mbox{\boldmath$\lambda$}}}
\makeatletter

\newcommand{\Rmnum}[1]{\expandafter\@slowromancap\romannumeral #1@}
\makeatother

\usepackage{multirow}
\usepackage{subfigure}

\begin{document}

\title{An empirical likelihood approach to reduce selection bias in voluntary samples} 

\author{Jae Kwang Kim \and Kosuke Morikawa} 
\date{} 
\maketitle 

\baselineskip .3in

\begin{abstract}

We address the weighting problem in voluntary samples under a  nonignorable sample selection model. Under the assumption that the sample selection model is correctly specified, we can compute a consistent estimator of the model parameter and construct the propensity score estimator of the population mean. We use the empirical likelihood method to construct the final weights for voluntary samples by incorporating the bias calibration constraints and the benchmarking constraints. Linearization variance estimation of the proposed method is developed. A limited simulation study is also performed to check the performance of the proposed methods. 
\end{abstract}

\newpage 

\section{ Introduction}

Probability sampling is a scientific tool for obtaining a representative sample from a target finite population.  Probability sampling allows constructing valid statistical inferences for finite population parameters.  Survey sampling is an area of statistics that deals with constructing efficient probability sampling designs and corresponding estimators. Classical approaches in survey sampling are discussed in \cite{cochran77},  \cite{sarndal92},  \cite{fuller09}, and \cite{tille2020}.  

 Despite the value of probability samples, non-probability samples are common even though an appropriate representation of the target population is not guaranteed. Collecting  a strict probability sample is almost impossible due to unavoidable issues such as frame undercoverage and nonresponse. However, statistical analysis of non-probability survey samples faces many challenges,  as documented by \cite{baker13}. Non-probability samples have unknown selection/inclusion mechanisms and are typically biased, and they do not represent the target population. A popular framework in dealing with biased non-probability samples is calibration weighting incorporating the auxiliary information observed throughout the finite population. Such calibration weighting method is based on the assumption that the selection mechanism for the non-probability sample is ignorable after adjusting for the auxiliary variables used for calibration weighting. Such
an assumption is essentially the missing at random (MAR) assumption  of \cite{rubin1976}. Using MAR assumptions, 
calibration weighting methods for non-probability samples have been discussed in \cite{dever2016} and \cite{elliott2017inference}, among others.

The MAR assumption, however, is a strong one. In many cases, the selection bias is not ignorable even after controlling on the auxiliary variables that are observed throughout the finite population. In that case, we need to build a model for the nonignorable sampling mechanism and develop an estimation procedure for the model parameters. 
Under a correctly specified model of the sampling mechanism, the generalized method of moments approach in \cite{kott10} and \cite{wang14} or the maximum likelihood estimation  approach considered in  \cite{pfeffermann11}, \cite{riddles15}, or \cite{morikawa2017} can be developed. See Chapter 8 of \cite{kimshao2021} for a comprehensive overview of statistical methods for nonignorable nonresponse models.

In this paper, motivated by \cite{qin2002}, we consider combining nonignorable selection model with empirical likelihood method to develop a unified approach to propensity score  weighting with calibration. The proposed method can handle nonignorable selection model and can incorporate the auxiliary variables through calibration weighting. Statistical inference with the final propensity score weighting is somewhat complicated due to the several steps in the final weighting.  While the idea of using a nonignorable response model to construct the final propensity weights for handling voluntary sample is natural, the literature on this research direction is somewhat limited. Thus, we present a systematic approach to using a nonignorable sampling model to adjust the selection bias and make inferences from a
non-probability sample. The empirical likelihood is nicely applied to create the final weights.  
The proposed method, however, is based on the model assumption for the sampling mechanism. If a single model assumption is not feasible, we can compute several estimates under different model scenarios and may consider sensitivity analysis \citep{copas01}.

The paper is organized as follows. In Section 2, the basic setup and the research problem are introduced. In Section 3, propensity score estimation method under nonignorable model is discussed. In Section 4, 
the final propensity score  weighting method using empirical likelihood is proposed and its variance estimation  is discussed. In Section 5, an illustrative example is used to present the computational detains of the proposed method. In Section 6, results from a limited simulation study are presented. An extension to a semiparametric nonignorable selection model is discussed in Section 7.  Some concluding remarks are made in Section 8. 

\section{Basic setup}

Consider a finite population of size $N$ and let $U=\{1,\cdots, N\}$ is the index set of the finite population. Let $S \subset U$ be the index set of sample.  
Let $\delta_i$ be the sample selection indicator of unit $i$ such that $\delta_i =1$ if $i \in S$ and $\delta_i=0$ otherwise.  
We observe $y_i$ only when $\delta_i=1$. We assume that the vectors of auxiliary variables $\bx_i$ are available  throughout the finite population. 
We are interested in estimating the population mean $\theta=N^{-1} \sum_{i=1}^N y_i$ from the sample. 

 Let $\pi (\bx,y) =P(\delta =1 \mid \bx, y)$ be the propensity score (PS) function for the sampling mechanism.  If $\pi(\bx,y)$ is  known, we can use the empirical likelihood (EL) method to estimate the parameter of interest. That is, we first find the maximizer of  
 $$ \ell( p) = \sum_{i \in S} \log (p_i ) 
$$ 
subject to $\sum_{i \in S} p_i =1$ and
\begin{equation} 
 \sum_{i \in S} p_i \pi (\bx_i, y_i) = W,
 \label{con3}
 \end{equation}
 where $W=P(\delta=1)$ is the marginal probability of $\delta=1$.   If $W=N^{-1} \sum_{i=1}^N \pi_i$ is unknown, we can estimate $W$ by the profile empirical likelihood method, as considered by \cite{qin2002}, or simply use $\widehat{W}_{HT}=n/N$, where $n=\left| S \right|$ is the realized sample size. Constraint (\ref{con3}) can be called the (internal) bias calibration constraint \citep{firth1998}.  
 In addition, we may impose 
\begin{equation} 
 \sum_{i \in S} p_i \bx_i = 
 \bar{\mathbf{X}}_N
 \label{calib}
 \end{equation} 
as an additional restriction, where $\bar{\mathbf{X}}_N = 
 N^{-1} \sum_{i=1}^N \bx_i $.  Constraint (\ref{calib}) can be called the benchmarking constraint.  Constraint (\ref{con3}) is used to remove the selection bias and constraint (\ref{calib}) is used to improve the efficiency of the resulting EL estimator. The implicit assumption for using (\ref{calib}) is that the outcome model is linear as in $E(Y \mid \bx) = \bx' \bbeta$ for some $\bbeta$.  If the outcome regression model is nonlinear, we can directly use the model calibration of \cite{wu2001model} to replace the benchmarking constraint in (\ref{calib}). 

Once $\hat{p}_i$ are obtained from the above optimization, the final estimator of $\theta$ is obtained by 
\begin{equation}
\hat{\theta}_{\rm EL}  = \sum_{i \in S} \hat{p}_i y_i .
\label{mele}
\end{equation}
The final estimator in (\ref{mele}) is often called the maximum empirical likelihood estimator (MELE) of $\theta$. The above empirical likelihood estimator has been considered by \cite{qin2002}, \cite{kim2009}, and \cite{berger2016}, among others.  
 
 If $W=N^{-1} \sum_{i=1}^N \pi_i$ is known, using the standard linearization, we can show that 
 \begin{equation}
\hat{\theta}_{\rm EL} = \frac{1}{N} \sum_{i=1}^N \left\{ \hat{y}_i + \frac{ \delta_i}{ \pi_i} \left( y_i - \hat{y}_i \right) \right\}  + o_p(n^{-1/2} ) ,
\label{result1}
 \end{equation}
 where $\hat{y}_i= \pi_i \hat{\beta}_1 + {\bx}_i' \hat{\bbeta}_2$  and 
 $$
 \begin{pmatrix}
\hat{\beta}_1 \\
\hat{\bbeta}_2 
\end{pmatrix}
= \left\{ 
\sum_{i \in S} \pi_i^{-2} \begin{pmatrix} 
 \pi_i -W \\
{\bx}_i- \bar{\mathbf{X}}_N 
\end{pmatrix} \begin{pmatrix} 
\pi_i -W \\
{\bx}_i- \bar{\mathbf{X}}_N 
\end{pmatrix}' \right\}^{-1}\sum_{i \in S} \pi_i^{-2} \begin{pmatrix} 
  \pi_i -W \\
{\bx}_i- \bar{\mathbf{X}}_N \end{pmatrix} y_i . $$
A sketched proof for (\ref{result1}) is presented in Appendix A. Result (\ref{result1}) means that the EL estimator is asymptotically equivalent to a regression estimator which satisfies the calibration constraints. In fact, the regression estimator in (\ref{result1}) is the design-optimal regression estimator of  \cite{rao1994} under Poisson sampling with negligible sampling fractions.

 The linearization formula in (\ref{result1}) is particularly useful in developing linearization variance estimation.  That is, we can use 
 \begin{equation}
 \hat{V}  = \frac{1}{N (N-1)} \sum_{i=1}^N \left( \hat{\eta}_i - \bar{\eta}_N \right)^2, 
 \label{varh}
 \end{equation}
 where $\hat{\eta}_i = \hat{y}_i + \delta_i \pi_i^{-1} \left( y_i - \hat{y}_i \right)$ and $\bar{\eta}_N=N^{-1} \sum_{i=1}^N \hat{\eta}_i$, to estimate the variance of $\hat{\theta}_{\rm EL}$.   
 The linearization variance estimator in (\ref{varh}) is derived based on the assumption that $\delta_i$ are mutually independent to each other. If they are correlated as in survey sampling, the joint probabilities of $(\delta_i, \delta_j)$ are needed to compute more accurate variance estimation. 

\section{Propensity score estimation under a parametric PS model} 

In probability sampling, $\pi_i$ are known and the  EL method in Section 2 can be directly applicable. In the voluntary sampling, we do not know the propensity score function $\pi(\bx,y)$. Instead, we may use a model for $\pi(\bx,y)$, say $\pi(\bx,y)=\pi(\bx,y; \phi_0)$ for some $\phi_0$, and develop methods for bias adjustment under the model. We first discuss how to estimate $\phi_0$ from the voluntary sample and then discuss estimation of $\theta=E(Y)$. 

  We assume that the propensity score function follows a parametric model such that 
    $\pi(\bx,y)=\pi(\bx,y; \phi_0)$ for some $\phi_0 \in \Phi \subset \mathbb{R}^q$. 
 The observed likelihood function of $\phi$ derived from the marginal density of $(\delta_i, \delta_i y_i )$ given $\bx_i$ can be written as 
    \begin{equation}
L_{\rm obs} ( \phi) = \prod_{i=1}^N \left\{ f(y_i \mid \bx_i) \pi(\bx_i, y_i ; \phi) \right\}^{\delta_i} \left\{ 1- \tilde{\pi} (\bx_i; \phi) \right\}^{1-\delta_i}
\label{obs1} 
    \end{equation}
where 
$$
\tilde{\pi}(\bx; \phi) = \mathbb{E} \left\{ \pi(\bx, Y; \phi) \mid \bx \right\}. 
$$
Thus, to construct the observed likelihood function in (\ref{obs1}), we may need to make a model assumption about $f( y \mid \bx)$.  Unfortunately, the resulting maximum likelihood estimator (MLE) of $\phi_0$ is not robust against model misspecification of the outcome model \citep{copas97}. 
Because we do not observe $y_i$ among $\delta_i=0$, we cannot directly apply the model diagnostic tools for complete response. 

Instead of using a model assumption for $f( y \mid \bx)$, we can make a model assumption for $f( y \mid \bx, \delta=1)$ with confidence, as $(\bx_i, y_i)$ are observed for $\delta_i=1$. Thus, we can safely assume that $f_1( y \mid \bx) \equiv f(y \mid \bx, \delta=1)$ is correctly specified. In this case, we can use the following identity which was originally proved by \cite{Pfeffermann1999}: 
 \begin{equation} \label{eq:pfeffermann}
    \tilde{\pi}(\bx;\phi) = \left[\int \{ \pi(\bx,y;\phi)\}^{-1} f_1( y \mid \bx) dy \right]^{-1}.
\end{equation}
Using (\ref{eq:pfeffermann}), we can construct 
\begin{equation} 
    \ell_{\rm obs}(\phi) = \sum_{i=1}^{N} \left\{ \delta_{i} \log {\pi}(\bx_{i}, y_i;\phi) + (1-\delta_{i}) \log \left( 1-\hat{\pi}(\bx_{i};\phi) \right) \right\},
    \label{obs2} 
\end{equation}
where 
\begin{equation}
\hat{\pi}(\bx;\bphi) = \left[\int \{ \pi(\bx,y;\bphi)\}^{-1} \hat{f}_1( y \mid \bx) dy \right]^{-1}
\label{hatpi}
\end{equation}
and $\hat{f}_1( y \mid \bx)$ is a consistent estimator of $f_1( y \mid \bx)$.  If we define 
$ \omega( \bx, y; \bphi) = \{ \pi(\bx, y; \bphi) \}^{-1} $
and $\widehat{\omega} (\bx; \bphi) = \{ \hat{\pi}(\bx; \bphi) \}^{-1} $, then (\ref{hatpi}) can be expressed as 
\begin{equation}
\widehat{\omega}(\bx;\bphi) = \int  \omega(\bx,y;\phi) \hat{f}_1( y \mid \bx) dy .
\label{hatpi2}
\end{equation}
 The propensity weight in (\ref{hatpi2}) can be called the smoothed propensity weight \citep{kim2013}. 
See Section 5 for some computational details of the smoothed propensity score function.   Furthermore, we impose  
\begin{equation}
\sum_{i=1}^N \frac{ \delta_i }{ \hat{\pi} (\bx_i; \phi) } = N 
\label{eq9}
\end{equation}
as a constraint for estimating the model parameter in the PS function. 
Including constraint (\ref{eq9}) into the ML estimation will improve the efficiency of the final estimation. It was originally proposed by \cite{cao2009improving}.

\cite{morikawa2017} show that the MLE can be obtained by solving 
\begin{equation} 
\sum_{i=1}^N \left\{  \frac{\delta_i}{\pi(\bx_i, y_i; \phi) } - 1 \right\} \mathbf{b}^* (\bx_i; \phi) = 0  ,
\label{7}
\end{equation}
where 
\begin{eqnarray*}
\mathbf{b}^*(\bx; \phi) 
&=& E\left\{   \mathbf{h}(\bx_i, Y; \phi) \pi(\bx_i, Y; \phi) \mid \bx_i, \delta_i=0\right\}, 
\end{eqnarray*}
$$
\mathbf{h} (\bx, y; \phi) = \frac{ \partial}{ \partial \phi} \mbox{logit} \{ \pi(\bx, y; \phi) \}, $$ and 
\begin{equation*}
 O(\bx, y; \phi) = \frac{ 1-\pi(\bx,y; \phi)}{ \pi(\bx,y; \phi) }. 
\end{equation*}

More generally, we can estimate $\phi$ by  solving 
\begin{equation}
 \hat{U}_{b} ( \phi) \equiv  \frac{1}{N} \sum_{i=1}^N \left\{ \frac{\delta_i}{ \pi(\bx_i, y_i ; \phi) } -1 \right\} \mathbf{b} (\bx_i; \phi) = 0
 \label{eq10}
 \end{equation}
 for some $\mathbf{b}(\bx; \phi)$ such that the solution exists uniquely. For example, we can use $\mathbf{b} (\bx_i; \phi ) = \bx_i$.   
 Parameter estimation using (\ref{eq10}) under a parametric PS  model was considered by \cite{kott10} and \cite{wang14}, among others.  
 Note that we can achieve (\ref{eq9}) by including the intercept term in $\mathbf{b} (\bx; \phi)$. If $\hat{\phi}$ is obtained from (\ref{eq10}), we can compute $\hat{\pi}_i = \pi(\bx_i, y_i; \hat{\phi})$ in the sample and we can estimate $\theta=E(Y)$ by 
\begin{equation}
\hat{\theta}_{\rm PS} = \frac{1}{N}\sum_{i \in S} \frac{1}{ \hat{\pi}_i} y_i .
\label{ps}
\end{equation}
Using the standard linearization, we can show that 
\begin{equation}
\hat{\theta}_{\rm PS} = \frac{1}{N} \sum_{i=1}^N \left\{ {\bgamma}' \mathbf{b}_i + \frac{ \delta_i}{ \pi_i} \left( y_i - {\bgamma}' \mathbf{b}_i \right) \right\}  + o_p(n^{-1/2} ) ,
\label{result2}
 \end{equation}
 where $\mathbf{b}_i = \mathbf{b} ( \bx_i; {\phi})$ is defined in (\ref{eq10}),  
 $$
  {\bgamma}' 
= E\left\{ Y \mathbf{h}_{0}(X,Y)'   \right\}
\left[ E\left\{  
\mathbf{b} (X)  \mathbf{h}_{0} (X, Y)'  \right\}\right]^{-1},$$
and $\mathbf{h}_{0}(X,Y) = \mathbf{h}(X,Y) \{1-\pi(X,Y)\}$. 
A sketched proof for (\ref{result2}) is presented in Appendix B.

The linearization formula in (\ref{result2}) can be used to construct the linearized variance estimation. We can apply the same linearization formula in (\ref{varh}) with a different formula for $\hat{\eta}_i$. Since we use a different linearization result in (\ref{result2}), we may use 
$$
\hat{\eta}_i = \hat{\mathbf{b}}_i' \hat{\bgamma} + \frac{\delta_i}{\hat{\pi}_i} \left( y_i - \hat{\mathbf{b}}_i' \hat{\bgamma} \right)$$
in applying the variance estimation formula (\ref{varh}), where 
$$
  \hat{\bgamma}' 
=\left( \sum_{i \in S}(\hat{\pi}_i^{-1}-1) \hat{\mathbf{h}}_{i}' y_i  \right)\left\{   
\sum_{i \in S} (\hat{\pi}_i^{-1} -1)  \hat{\mathbf{b}}_{i}  \hat{\mathbf{h}}_{i}'   \right\}^{-1}$$
and 
$\phi$ is replaced by $\hat{\phi}$ in the linearization formula. 

Now, to incorporate the auxiliary information, we consider the class of the  regression PS estimator 
\begin{equation}
\hat{\theta}_{RPS} ({\bbeta}) = \frac{1}{N} \sum_{i=1}^N 
\left\{ \bx_i' {\bbeta} + \frac{ \delta_i}{\hat{\pi}_i } \left( y_i - \bx_i' \bbeta \right) \right\}
\label{gps}
\end{equation}
where $\bbeta$ is to be determined. 

If we choose $\hat{\bbeta}^*$ such that the asymptotic variance of $\hat{\theta}_{RPS} ({\bbeta})$ is minimized among the class in (\ref{gps}), we obtain the optimal PS estimator.
The optimal $\hat{\bbeta}^*$ can be written as 
$$ 
\hat{\bbeta}^* 
= \left[ 
\widehat{V} \left\{ 
\sum_{i \in S} \hat{\pi}_i^{-1} \bx_i 
\right\} \right]^{-1} \widehat{Cov}\left\{ 
\sum_{i \in S} \hat{\pi}_i^{-1} \bx_i , \sum_{i \in S} \hat{\pi}_i^{-1} 
 y_i \right\}$$
 and we can use Taylor expansion to derive the explicit formula for the variance-covariance matrix. When $\hat{\phi}$ is obtained from (\ref{eq10}), we can compute the optimal estimator by solving 
 $$ 
 \sum_{i \in S} \hat{\pi}_i^{-1} (1- \hat{\pi}_i )\left( y_i - \bx_i' \bbeta - \mathbf{b}_i' \bgamma \right) \bx_i = 0   
 $$
 and 
 $$
 \sum_{i \in S} \hat{\pi}_i^{-1} (1- \hat{\pi}_i )\left( y_i - \bx_i' \bbeta - \mathbf{b}_i' \bgamma \right) \hat{\mathbf{h}}_i = 0   
 $$
 for $\bbeta$ and $\bgamma$.  \cite{morikawa2018} 
 proposed an adaptive optimal estimator achieving the semiparametric lower bound. 

\section{Empirical likelihood method}

Using the parameter estimation in Section 3, we can  apply the EL method for final PS weighting.  That is, we first find the maximizer of  
 $$ \ell( p) = \sum_{i \in S} \log (p_i ) 
$$ 
subject to $\sum_{i \in S} p_i =1$, 
\begin{equation} 
 \sum_{i \in S} p_i \pi (\bx_i, y_i; \hat{\phi}) = N^{-1} \sum_{i=1}^N \hat{\pi} (\bx_i; \hat{\phi}),
 \label{con6}
 \end{equation}
 and 
 \begin{equation}
 \sum_{i \in S} p_i  \bx_i = N^{-1} \sum_{i=1}^N \bx_i,  
 \label{con7}
 \end{equation}
 where $\hat{\pi} (\bx_i; \hat{\phi}) $ is defined in (\ref{hatpi}) and $\hat{\phi}$ is computed from (\ref{eq10}).  Condition (\ref{con6}) is the bias calibration condition.

 Note that the final weights are obtained in two steps. In the first step, a consistent estimator $\hat{\phi}$ of the model parameter in the PS model $\pi(\bx, y; \phi)$ is computed. In the second step, we treat $\hat{\pi}_i=\pi(\bx_i, y_i; \hat{\phi})$ as if the true inclusion probability and apply the EL method incorporating the bias-calibration constraints and the calibration constraint.

 Now, to discuss the asymptotic property of the final EL estimator $\hat{\theta}_{\rm EL}$, we apply the same two-step procedure to obtain Taylor linearization.  In the first step, ignoring the uncertainty in $\hat{\phi}$ for now, we can apply the linearization method for obtaining (\ref{result1}) to get 
 
  \begin{equation}
\hat{\theta}_{\rm EL} = \frac{1}{N} \sum_{i=1}^N \left\{ \hat{y}_i^{(0)} + \frac{ \delta_i}{ \pi (\bx_i, y_i ; \hat{\phi})} \left( y_i - \hat{y}_i^{(1)} \right) \right\}  + o_p(n^{-1/2} ) ,
\label{result3}
 \end{equation}
where  $\hat{y}_i^{(0)}  = \hat{\pi} (\bx_i; \hat{\phi} ) \hat{\beta}_1 + \tilde{\bx}_i' \hat{\bbeta}_2 $, $\hat{y}_i^{(1)}  = {\pi} (\bx_i, y_i; \hat{\phi} ) \hat{\beta}_1 + {\bx}_i' \hat{\bbeta}_2 $ and 
 $$
 \begin{pmatrix}
\hat{\beta}_1 \\
\hat{\bbeta}_2 \\
\end{pmatrix}
= \left\{ 
\sum_{i \in S} \hat{\pi}_i^{-2} \begin{pmatrix} 
\hat{\pi}_i - \hat{W} \\
{\bx}_i - \bar{\mathbf{X}}_N  \end{pmatrix} \begin{pmatrix} 
\hat{\pi}_i - \hat{W} \\
{\bx}_i - \bar{\mathbf{X}}_N
\end{pmatrix}' \right\}^{-1}\sum_{i \in S} \hat{\pi}_i^{-2} \begin{pmatrix} 
\hat{\pi}_i - \hat{W} \\
{\bx}_i - \bar{\mathbf{X}}_N
\end{pmatrix} y_i . $$ 
 Here, $\hat{\pi}_i = \pi(\bx_i, y_i ; \hat{\phi})$ and $\hat{W}=N^{-1} \sum_{i=1}^N{\pi} (\bx_i; \hat{\phi} ) $.


Note that we can express (\ref{result3}) as 
 \begin{equation}
\hat{\theta}_{\rm EL} = \frac{1}{N} \sum_{i=1}^N \left\{ \hat{\beta}_1 \left( \hat{\pi}(\bx_i; \hat{\phi} ) - \delta_i\right)  + {\bx}_i' \hat{\bbeta}_2 +  \frac{ \delta_i}{\hat{\pi}_i} \left( y_i - {\bx}_i' \hat{\bbeta}_2 \right) \right\} + 
o_p(n^{-1/2} ) .
\label{result4}
 \end{equation}
In the second step, we need to take into account the uncertainty in $\hat{\phi}$. To do this, we apply the Taylor expansion with respect to $\phi$ and obtain the final influence function. 

Now, to apply the Taylor expansion with respect to $\phi$, define 
 \begin{eqnarray*}
 \hat{\theta}_{\ell} ( \hat{\phi}) &=& \frac{1}{N} \sum_{i=1}^N  \hat{\beta}_1 \left( \hat{\pi}(\bx_i; \hat{\phi} ) - \delta_i\right)  + \frac{1}{N} \sum_{i=1}^N \left\{ {\bx}_i' \hat{\bbeta}_2 +  \frac{ \delta_i}{\hat{\pi}_i} \left( y_i - {\bx}_i' \hat{\bbeta}_2 \right) \right\} \\
 &:=& \hat{\theta}_{\ell, 1} ( \hat{\phi}) + \hat{\theta}_{\ell, 2} (\hat{\phi}) 
 \end{eqnarray*} 
 and note that (\ref{result3}) can be written as 
 $$ \hat{\theta}_{\rm EL} = \hat{\theta}_{\ell} ( \hat{\phi} ) +o_p (n^{-1/2}). $$
 
 We can apply Taylor linearization to $\hat{\theta}_{\ell, 1}( \hat{\phi})$ with respect to $\phi$ to get 
 \begin{eqnarray}
\hat{\theta}_{\ell, 1} ( \hat{\phi} ) &=& \frac{1}{N} \sum_{i=1}^N  \hat{\beta}_1 \left( \hat{\pi}(\bx_i; {\phi} ) - \delta_i\right)  
+ E\left\{ N^{-1} \sum_{i=1}^N \hat{\beta}_1 \frac{\partial}{ \partial \phi'} \hat{\pi} (\bx_i ; \phi) \right\} \left( \hat{\phi} - \phi \right) + o_p(n^{-1/2}) \notag \\
&=&  \frac{1}{N} \sum_{i=1}^N  \left[ \hat{\beta}_1 \{ \hat{\pi} (\bx_i; \phi) + \bkappa_1' \mathbf{b}_i \} - \frac{ \delta_i}{ \pi_i} \hat{\beta}_1 \left\{  \pi_i+  \bkappa_1' \mathbf{b}_i \right\}\right]  + o_p( n^{-1/2} ), \label{res-1}
\end{eqnarray}
where $\pi_i = \pi(\bx_i, y_i; \phi)$,  $\bkappa_1' = - E\left\{  \hat{\mathbf{g}}_1 (X ; \phi)'  \right\} \left[  E \left\{  \mathbf{b} (X) \mathbf{h}_0 (X,Y)' \right\} \right]^{-1}  
$
and 
\begin{eqnarray}
\hat{\mathbf{g}}_1 (\bx ; \phi) &=&  \frac{\partial }{\partial \bphi} \hat{\pi}( \bx; \bphi) \notag \\
&=& \{ \hat{\pi} (\bx; \bphi) \}^2
\left[\int \left\{ \frac{1}{\pi(\bx,y; \bphi)} -1\right\} \mathbf{h}(\bx, y; \phi) \hat{f}_1( y \mid \bx) dy\right].   \label{h1}
\end{eqnarray}

Also, we can apply the Taylor linearization to $\hat{\theta}_{\ell, 2} ( \hat{\phi})$ with respect to $\phi$ to get 
\begin{eqnarray}
\hat{\theta}_{\ell, 2} ( \hat{\phi} ) &=& \frac{1}{N} \sum_{i=1}^N \left\{ {\bx}_i' \hat{\bbeta}_2 +  \frac{ \delta_i}{{\pi}_i} \left( y_i - {\bx}_i' \hat{\bbeta}_2 \right) \right\}  
\notag 
\\
&& - 
E\left\{ N^{-1} \sum_{i=1}^N \delta_i \pi_i^{-2} \pi_i (1- \pi_i) (y_i - {\bx}_i' \hat{\bbeta}_2 ) \mathbf{h}_i' \right\} \left( \hat{\phi} - \phi \right) + o_p(n^{-1/2}) 
\notag \\
&=&\frac{1}{N} \sum_{i=1}^N \left\{ {\bx}_i' \hat{\bbeta}_2 + \mathbf{b}_i' \bkappa_2 +  \frac{ \delta_i}{{\pi}_i} \left( y_i - {\bx}_i' \hat{\bbeta}_2- \mathbf{b}_i' \bkappa_2  \right) \right\}  + o_p(n^{-1/2}) ,\label{res-2}
\end{eqnarray}
where 
$$\bkappa_2' =   E\left\{   (Y- \tilde{\bx}' \hat{\bbeta}_2) \mathbf{h}_0 (X,Y)' \right\} \left[  E \left\{ \mathbf{b} (X) \mathbf{h}_0 (X,Y)' \right\} \right]^{-1} 
$$
and $\mathbf{h}_0 (X,Y)= \mathbf{h} ( X, Y) \{ 1- \pi(X,Y) \}$. 

Combining (\ref{res-1}) and (\ref{res-2}), we obtain 
 \begin{eqnarray}
\hat{\theta}_{\rm EL} &=& \frac{1}{N} \sum_{i=1}^N  \left\{  \hat{y}_i^{(0)} + \frac{ \delta_i}{ \pi_i}  \left(  y_i - \hat{y}_i^{(1)}  \right)\right\} + o_p(n^{-1/2}) ,
\label{linear} 
\end{eqnarray} 
where 
$$\hat{y}_i^{(0)}  =
\hat{\beta}_1 \{ \hat{\pi} (\bx_i; \phi) + \mathbf{b}_i' \bkappa_1\} + {\bx}_i' \hat{\bbeta}_2 + \mathbf{b}_i' \bkappa_2 $$
and 
$$\hat{y}_i^{(1)}  =
\hat{\beta}_1 \left( \hat{\pi}_i + \mathbf{b}_i' \bkappa_1\right) + {\bx}_i' \hat{\bbeta}_2 + \mathbf{b}_i' \bkappa_2 . $$
The linearization formula (\ref{linear}) shows that the EL estimator is asymptotically equivalent to a version of regression estimator but we use $\hat{\pi}(\bx_i; \phi)$ instead of ${\pi}_i=\pi(\bx_i; y_i ; {\phi})$ in $\hat{y}_i^{(0)}$ because $y_i$ are observed only in the sample. The uncertainty associated with $\hat{\bbeta}$ is asymptotically negligible. 

The linearization formula can be used to develop the variance estimation for $\hat{\theta}_{\rm EL}$.  We can use 
\begin{equation}
\hat{\bkappa}_1'= - \sum_{i=1}^N \hat{\mathbf{g}}_1 (\bx_i ; \hat{\phi} )' \left\{ \sum_{i=1}^N \delta_i  (\hat{\pi}_i^{-1} -1)  \hat{\mathbf{b}}_i \hat{\mathbf{h}}_i'  \right\}^{-1}   
\label{kappa1}
\end{equation}
and 
\begin{equation}
\hat{\bkappa}_2' = \left\{  \sum_{i=1}^N \delta_i \left( \hat{\pi}_i^{-1} -1 \right) \left(y_i  - {\mathbf{x}}_i' \hat{\bbeta}_2 \right) \hat{\mathbf{h}}_i' \right\} \left\{ \sum_{i=1}^N \delta_i \left( \hat{\pi}_i^{-1} -1 \right) \hat{\mathbf{b}}_i \hat{\mathbf{h}}_i'  \right\}^{-1}    
\label{kappa2}
\end{equation}
to compute the linearization variance estimator. 

\section{An illustrative example}

We use a toy example to demonstrate the method and describe the computational details. Suppose that we have two auxiliary variables, $X_1$ and $X_2$, and the PS model is given by 
\begin{equation}
P( \delta = 1 \mid x_1, x_2, y ) = \frac{\exp ( \phi_0 + \phi_1 x_1 + \phi_2 y) }{ 1+ \exp  ( \phi_0 + \phi_1 x_1 + \phi_2 y ) }:= \pi(x_1, y; \bphi). 
\label{logistic}
\end{equation}
 Parameter $\bphi=(\phi_0, \phi_1, \phi_2)'$ is estimated by solving 
\begin{equation}
\hat{U}_b ( \bphi) \equiv \frac{1}{N} \sum_{i=1}^N \left\{ \frac{ \delta_i}{ \pi(x_{1i}, y_i ; \bphi) } -1 \right\} \mathbf{b}_i  =0 
\label{ee3}
\end{equation}
where $\mathbf{b}_i= \mathbf{b}( \bx_i)$ is a vector such that the solution to (\ref{ee3}) exists almost everywhere. 

The Newton method for solving (\ref{ee3}) can be written as 
\begin{equation}
\hat{\bphi}^{(t+1)} = \hat{\bphi}^{(t)} + \left\{
\sum_{i=1}^N \delta_i  O_i^{(t)} \mathbf{b}_i \mathbf{h}_i'  \right\}^{-1} \sum_{i=1}^N \left\{ \frac{ \delta_i}{ \pi(x_{1i}, y_i ; \hat{\bphi}^{(t)}) } -1 \right\}  \mathbf{b}_i 
\label{newton}
\end{equation}
where $O_i^{(t)} = \{ \pi(x_{1i}, y_i; \hat{\bphi}^{(t)} )\}^{-1} -1= \exp ( - \hat{\phi}_0^{(t)} -  \hat{\phi}_1^{(t)}  x_{1i} - \hat{\phi}_2^{(t)} y_i )$ and  $\mathbf{h}_i = (1, x_{1i}, y_i)'$. 
Because $\left\{
\sum_{i=1}^N \delta_i  O_i^{(t)} \mathbf{b}_i \mathbf{h}_i'  \right\}$ is not symmetric, the computation for 
(\ref{newton}) can be unstable.  

One way to avoid the computational problem is to make the parameter estimation problem an optimization problem. One way is to find $\hat{\bphi}$ by finding the minimizer of 
$$ Q( \bphi) = \hat{U}_b ( \bphi)' \hat{U}_b ( \bphi) .
$$
In this case, the Newton method can be expressed by 
\begin{equation}
\hat{\bphi}^{(t+1)} = \hat{\bphi}^{(t)} -  \left\{
\dot{U}_b ( \hat{\bphi}^{(t)} )'\dot{U}_b ( \hat{\bphi}^{(t)} )\right\}^{-1} \dot{U}_b ( \hat{\bphi}^{(t)} )' \hat{U}_b( \hat{\bphi}^{(t)} )
\label{newton2}
\end{equation}
where 
$$\dot{U}_b ( \hat{\bphi}^{(t)} )' = - N^{-1} \sum_{i=1}^N \delta_i  O_i^{(t)} \mathbf{b}_i \mathbf{h}_i' . $$

Now, let us discuss how to compute the smoothed propensity score function $\hat{\pi}(\bx;\phi)$ in (\ref{hatpi}). 
Since $\hat{\bphi}$ is obtained by solving (\ref{ee3}), it satisfies 
$$ \sum_{i=1}^N \delta_i \omega(\bx_i, y_i ; \hat{\bphi}) \mathbf{b} (\bx_i) = \sum_{i=1}^N \mathbf{b} (\bx_i)  . 
$$
Thus, by (\ref{hatpi2}), the smoothed weight $\widehat{\omega} ( \bx; \hat{\phi})= \int\omega(\bx, y ; \hat{\bphi}) \hat{f}_1( y \mid \bx) dy$ should also satisfy 
\begin{equation}
\sum_{i=1}^N \delta_i \widehat{\omega}(\bx_i ; \hat{\bphi}) \mathbf{b} (\bx_i) = \sum_{i=1}^N \mathbf{b} (\bx_i)
\label{calib2}
\end{equation}
which is the calibration equation for $\mathbf{b}(\bx)$. That is, the smoothed weights should satisfy the same calibration equation as the original weights. 

Since $\omega(\bx, y; \bphi)=1+ \exp ( - \phi_0 - \phi_1 x_1 - \phi_2 y) $, one way to achieve (\ref{calib2}) easily is to use 
\begin{equation}
\hat{f}_1 ( y \mid \bx) = \frac{1}{ \sqrt{2\pi}\hat{\sigma} } \exp \left\{  - 
\frac{1}{2\hat{\sigma}^2} \left( y - \bx' \hat{\balpha} \right)^2 \right\}
\label{gaussian} 
\end{equation}
for some $\hat{\balpha}$ and $\hat{\sigma}^2$.  We can use the moment-generating function formula of Gaussian distribution to compute 
\begin{equation}
\int \exp \left(  - \phi_2 y \right) \hat{f}_1 \left( y \mid \bx \right) d y = \exp \left( - \phi_2 \bx' \hat{\balpha}+ \frac{1}{2} \phi_2^2 \hat{\sigma}^2  \right). 
\label{35}
\end{equation}
We can use constraint (\ref{calib2}) to compute $\hat{\balpha}$ and $\hat{\sigma}^2$. That is, $\hat{\balpha}$ is the solution to the following estimating equation: 
\begin{equation}
\sum_{i=1}^N  \delta_i \left\{ 1 + \exp \left( - \hat{\phi}_0 - \hat{\phi}_1 x_{1i} - \hat{\phi}_2 \bx_i'{\balpha} + \frac{1}{2} \hat{\phi}_2^2\hat{\sigma}^2\right) \right\}\mathbf{b}_i = \sum_{i=1}^N (1- \delta_i) \mathbf{b}_i.
\label{calib3}
\end{equation}
To obtain the unique solution, we may set $\hat{\sigma}^2=1$. 
Once $\hat{\balpha}$ is computed from (\ref{calib3}), we can obtain 
$$\widehat{\pi} (\bx_i; \hat{\bphi})
= \left\{ 1+ \exp \left( - \hat{\phi}_0 - \hat{\phi}_1 x_{1i} - \hat{\phi}_2 \bx_i'\hat{\balpha} + \frac{1}{2} \hat{\phi}_2^2\hat{\sigma}^2\right) \right\}^{-1}.$$

 Once we obtain $\hat{\pi}( \bx; \hat{\bphi})$, we can apply the EL method to obtain the final PS weights. The actual computation for the EL weighting can be implemented using the method of \cite{chen2002}.
 Once $\hat{p}_i$ are obtained by the optimization problem using the bias calibration constraint (\ref{con6}) and the balancing constraint (\ref{calib}), we can construct the maximum EL estimator of $\theta$ by $\hat{\theta}_{\rm EL} = \sum_{i \in S} \hat{p}_i y_i$. 

To compute the linearization variance estimator, we need to compute $\hat{\bkappa}_1$ and $\hat{\bkappa}_2$ in (\ref{kappa1}) and (\ref{kappa2}), respectively. To compute $\hat{\bkappa}_1$, we need to compute $\hat{\mathbf{g}} (\bx ; \phi)$ in (\ref{h1}). Since $\mathbf{h}(\bx, y) = (1, x_1, y)'$, we can express  
$$ \hat{\mathbf{g}} (\bx; \phi) = \{ \hat{\pi} (\bx; \bphi) \}^2 \left\{ 
\int \omega(\bx, y; \phi) (1, x_1, y)' \hat{f}_1 (y \mid \bx) d y -1 \right\} := (\hat{g}_1, \hat{g}_2, \hat{g}_3 )'.$$
Since 
$\hat{\pi} (\bx; \phi)$ satisfies (\ref{calib2}), we can obtain 
\begin{eqnarray*}
\hat{g}_1 &=& \{ \hat{\pi} (\bx; \bphi) \}^2 \left\{ 
\int \omega(\bx, y; \phi) \hat{f}_1 (y \mid \bx) d y -1 \right\} \\
&=& \{ \hat{\pi} (\bx; \bphi) \}^2\left\{ \frac{1}{\hat{\pi} (\bx; \bphi)} - 1 \right\} \\
&=& \hat{\pi} (\bx; \bphi) \left\{ 1- {\hat{\pi} (\bx; \bphi)}  \right\}\end{eqnarray*}
and 
$$ \hat{g}_2 = x_{1} \cdot  \hat{\pi} (\bx; \bphi) \left\{ 1- {\hat{\pi} (\bx; \bphi)}  \right\}.$$
Now, to compute $\hat{g}_3$, note that 
\begin{eqnarray*}
\hat{g}_3 &=& \{ \hat{\pi} (\bx; \bphi) \}^2 
\int \exp (- \phi_0 - \phi_1 x_1 - \phi_2 y) y  \hat{f}_1 (y \mid \bx) d y  \\
&=&  \{ \hat{\pi} (\bx; \bphi) \}^2 
 \exp (- \phi_0 - \phi_1 x_1)  \int \exp (- \phi_2 y) y  \hat{f}_1 (y \mid \bx) d y . \end{eqnarray*} 
 If $y \mid (\bx, \delta =1) \sim N( \mathbf{x}' \hat{\balpha}, \hat{\sigma}^2)$, then we can obtain
 \begin{eqnarray*}
 \hat{g}_3 &=&  \{ \hat{\pi} (\bx; \bphi) \}^2 
 \exp (- \phi_0 - \phi_1 x_1)  
  \int \exp (- \phi_2 y) y \cdot \frac{1}{ \sqrt{2 \pi} \hat{\sigma}} \exp \left\{- \frac{1}{2 \sigma^2} ( y - \bx'\hat{\balpha} )^2 \right\} dy \\
  &=& \{ \hat{\pi} (\bx; \bphi) \}^2 
 \exp (- \phi_0 - \phi_1 x_1)  
  \int  y \cdot \frac{1}{ \sqrt{2 \pi} \hat{\sigma}} \exp \left\{- \frac{1}{2 \hat{\sigma}^2} ( y - \bx'\hat{\balpha} + \hat{\sigma}^2 \phi_2)^2 \right\} dy \\
  & & \times \exp \left\{ - \frac{1}{ \hat{\sigma}^2} (\bx' \hat{\balpha})^2 + \frac{1}{2 \hat{\sigma}^2} (\bx' \hat{\balpha} - \hat{\sigma}^2 \phi_2)^2\right\}
  \\
  &=&
\{ \hat{\pi} (\bx; \bphi) \}^2 
 \exp (- \phi_0 - \phi_1 x_1) \exp \left( \frac{1}{2} \phi_2^2 \hat{\sigma}^2 - (\bx' \hat{\balpha}) \phi_2 \right) \cdot  \left( \bx' \hat{\balpha} - \phi_2 \hat{\sigma}^2  \right) \\
&=& \left( \bx' \hat{\balpha} - \phi_2 \hat{\sigma}^2  \right)\cdot  \hat{\pi} (\bx; \bphi) \left\{ 1- {\hat{\pi} (\bx; \bphi)}  \right\},
\end{eqnarray*}
where the last equality follows from (\ref{35}). 

\section{Simulation Study}

To test our theory, we performed a limited simulation study. We consider two outcome models for the simulation study: 
\begin{enumerate}
    \item M1:  $y_i = -4 + x_{1i} +   x_{2i} + e_i$
    \item M2: $y_i = 0.5*(x_{1i}+ x_{2i}-5)^2 -1.5 + e_i$
\end{enumerate}
We use $x_{1i}, x_{2i} \sim N(2,1)$ and $e_i \sim N(0,1)$. 
Regarding the response mechanism, we used $\delta_i \sim \mbox{Bernoulli} (\pi_i)$ where 
$$
\pi_i = \frac{ \exp ( \phi_0 +  \phi_1 x_{1i} + \phi_2 y_i)}{ 1+\exp ( \phi_0 + \phi_1 x_{1i} + \phi_2 y_i)}
$$
where $(\phi_0, \phi_1, \phi_2)= (-2, 1.0, 0.5)$.  
We generated a sample of $(x_{1i}, x_{2i}, y_i, \delta_i)$ from the above mechanism with sample size $N=5,000$. The overall sampling rate is 50\% in both scenarios. 
We repeat the Monte Carlo sampling independently $B=1,000$ times.

From each sample, 
we computed four estimators.
\begin{enumerate}
\item (EL-MAR) The EL estimator assuming 
$$
\pi (\bx, y; \phi)  = \frac{ \exp ( \phi_0 + \phi_1 x_{1i} + \phi_2 x_{2i} )}{ 1+\exp ( \phi_0 + \phi_1 x_{1i} + \phi_2 x_{2i})}
$$
as the response model. Note that the response model is MAR and incorrectly specified. 

\item (PS) The PS estimator in (\ref{ps}) under the correct model.  
The parameter $\bphi$ for the PS model is estimated by solving 
$$ \sum_{i=1}^N \left( \frac{ \delta_i}{ \pi(x_{1i}, y_i; \bphi) } -1 \right) \left( \begin{array}{l}
1 \\ \bx_i  \end{array} \right) =\left( \begin{array}{l}
0 \\ 0  \end{array} \right) $$
where $\bx_i = (x_{1i}, x_{2i})'$. 
\item (EL-1) The maximum EL estimator using the estimated $\pi_i$ in the correct model with the bias calibration condition in (\ref{con6}). 
\item (EL-2) The maximum EL estimator using estimated $\pi_i$ under the correct model and satisfying the calibration constraint (\ref{calib}) in addition to (\ref{con6}). 

\end{enumerate}
 The simulation results for the point estimators are summarized in Table 1. The performance is as expected. The EL method assuming the MAR model is severely biased. The EL method improves the efficiency over the PS method under the correct model, but the efficiency gain is not high in this simulation setup.  
 
 In addition to point estimators, we also calculated the normal-based interval estimator for the proposed EL-2 estimator with 95\% nominal coverage rate. The realized coverage rates are 95\% and 96\% for Scenarios M1 and M2, respectively. Under M2, the linearized variance estimator is slightly overestimated with a relative bias equal to 8\% approximately.   
 The slight overestimation of variance for M2 seems to come from the fact that the linear regression model $\hat{f}_1( y \mid \bx)$ given by (\ref{gaussian}) in computing $\hat{\pi}( \bx ; \phi)$ through (\ref{hatpi}) is incorrectly specified under M2. 
 Under M1, the normal model assumption for $\hat{f}_1( y \mid \bx)$ is roughly satisfied and the linearization variance estimator is nearly unbiased.  


\begin{table}[ht]
\centering
\caption{Monte Carlo biases, variance, and MSE of the estimators computed from 1,000 MC samples}
\begin{tabular}{c|cccc}
  \hline
Scenario & Method & Bias & Var($\times 1000$) & MSE($\times 1000$) \\ 
  \hline
  \multirow{5}{*}{M1} & Full & 0.00 & 0.55 & 0.55 \\
&   EL (MAR) &0.25  & 1.34 &  61.11 \\
 & PS & 0.00 & 2.09 & 2.11  \\
 & EL-1  & 0.00 & 1.94 & 1.95 \\
 & EL-2 & 0.01 & 2.03 & 2.08 \\
   \hline
 \multirow{5}{*}{M2}   & Full & 0.00 & 0.93 &0.93 \\
& EL (MAR) & 0.62   & 2.24 & 386.14 \\
  & PS & 0.00 & 2.75 & 2.75\\
 & EL-1  & 0.00 & 2.73  & 2.73 \\
 & EL-2 &  0.00 &  2.72 & 2.72 \\ \hline
\end{tabular}
\end{table}

\section{Extension to a semiparametric PS  model}

In this section, we briefly present ideas for an extension to the semiparametric propensity score model. The parametric model approach presented in Section 3 is easy to derive the theory but may be subject to the bias due to model mis-specification. To resolve the problem, we can consider a more flexible propensity score model. One possible model is   
\begin{equation}
\pi(\bx_i, y_i) \equiv Pr \left( \delta_i=1 \mid \bx_i, y_i \right)  = \frac{\exp \left\{  g(\mathbf{x}_i)  + \phi y_i  \right\} }{ 1+ \exp \left\{ g( \mathbf{x}_i)
 + \phi y_i
 \right\} }, \label{6-4-7}
\end{equation}
where $g( \mathbf{x})$ is completely unspecified. The semiparametric PS model in (\ref{6-4-7}) is first considered by \cite{kimyu2011} and further discussed by \cite{franks2022}. Model (\ref{6-4-7}) implies that  
\begin{equation}
f_0 \left( y_i \mid {\bx}_i \right)= f_1 \left( y_i \mid
{\bx}_i \right) \times \frac{\exp \left( \gamma y_i \right) }{
E \left\{  \exp \left( \gamma Y \right) \mid {\bx}_i, \delta_i=1
\right\}}, \label{6-4-8}
\end{equation}
where $\gamma=-\phi$ and 
$f_\delta \left( y_i \mid \bx_i \right) = f \left( y_i \mid \bx_i, \delta_i = \delta \right)$.


Under this model, we can use 
$$ E\left\{ \frac{ \delta}{ \pi (\bx, y) } -1 \mid \mathbf{x} \right\} = 0  $$
to obtain
$$ \exp\{ g(\bx) \} = \frac{  E\{ \delta \exp ( \gamma y) \mid \bx\} }{ E\{  1-\delta \mid \bx \} } . $$
 For known $\gamma$ case, we can use kernel regression estimator
$$ \exp \{ \hat{g}_{\gamma} (\bx) \} = \frac{ \sum_{i=1}^n  \delta_i \exp ( \gamma y_i) K_h ( x - x_i) }{ \sum_{i=1}^n (1-\delta_i) K_h (x - x_i ) }  $$
to obtain the following profile PS function 
\begin{equation}
\hat{\pi}_p (\bx_i, y_i; \gamma ) = \frac{ \exp\{ \hat{g}_{\gamma} (\bx_i) - \gamma y_i  \} }{ 1+ \exp\{ \hat{g}_{\gamma}  (\bx_i) - \gamma y_i  \} } .  
\label{prof}
\end{equation}

For estimation of $\gamma$, \cite{shao2016} suggested using the GMM method based on some moment conditions.  We can use the profile log likelihood to find the maximum likelihood estimator of $\gamma$: 
\begin{equation}
\ell_{p}( \gamma) = \sum_{i=1}^N \left[ \delta_i \log\{   \hat{\pi}_p (\bx_i, y_i; \gamma ) \} 
+ (1- \delta_i ) \log\{  1- \tilde{\pi}_p (\bx_i; \gamma ) \} \right] 
\label{obs3}
\end{equation}
where 
\begin{equation*} 
    \tilde{\pi}_p(\bx;\gamma) = \left[\int \{ \hat{\pi}_p(\bx,y;\gamma)\}^{-1} \hat{f}_1( y \mid \bx) dy \right]^{-1}.
\end{equation*}
See \cite{uehara2020} for more details of the profile maximum likelihood estimator of $\gamma$. 
Once $\hat{\gamma}$ is obtained by finding the maximizer of $\ell_p( \gamma)$ in (\ref{obs3}), we can apply the same EL method to find the final weights. That is, we first find the maximizer of  
 $$ \ell( p) = \sum_{i \in S} \log (p_i ) 
$$ 
subject to $\sum_{i \in S} p_i =1$, 
\begin{equation*} 
 \sum_{i \in S} p_i \hat{\pi}_p (\bx_i, y_i; \hat{\gamma}) = N^{-1} \sum_{i=1}^N \tilde{\pi}_p (\bx_i; \hat{\gamma}),
 \end{equation*}
 and (\ref{calib}). Investigating the asymptotic properties of the resulting EL estimator is beyond the scope of the paper and will be presented elsewhere. 

\section{Concluding Remarks}

We have developed an EL-based approach to propensity score estimation with an unknown propensity score function. Under the assumption that the PS model is correctly specified, we can obtain a consistent estimator of the PS model parameters and construct the final EL weights. The final EL weights can incorporate the calibration constraints in addition to the bias correction constraint. The two-step linearization method described in Section 4 can be used to develop a linearized variance estimator of the maximum EL estimator of the population mean. 
If the PS model is unknown, we may consider a more flexible model, as discussed in Section 7.

There are several possible extensions. Instead of using a single PS model, we can consider multiple PS models and employ multiple constraints for bias correction in the final estimation. This is similar in spirit to multiple robust estimation of \cite{han2013} and \cite{chenhaziza2017}. Also, the proposed method can be extended to data integration problems \citep{yang2020}, combining a voluntary sample with a probability sample. Such extensions will be presented elsewhere.

\section*{Acknowledgements}

The authors thank the guest editor, Professor Partha Lahiri, for invitation to the special issue and for the constructive comments.  Research by the first author was partially supported by  a grant from the Iowa Agriculture and
Home Economics Experiment Station, Ames, Iowa.
Research by the second author was supported by JST CREST Grant Number JPMJCR1763 and MEXT Project for Seismology toward Research Innovation with Data of Earthquake (STAR-E) Grant Number JPJ010217.
\newpage 
\section*{Appendix}

\subsection*{A. Proof of (\ref{result1})}

The EL optimization problem can be expressed as maximizing $\ell (\mathbf{p}) = \sum_{i \in S} \log(p_i)$ subject to $\sum_{i \in S} p_i=1$, bias calibration constraint (\ref{con3}), and the benchmarking constraint (\ref{calib}). To incorporate the three constraints, the EL weights can be written as 
$$
\hat{p}_i = \frac{1}{n} \frac{1}{1+ \hat{\lambda}_1 (\pi_i-W )  + \hat{\blambda}_2' (\bx_i-\bar{\mathbf{X}}_N) } := \hat{p}_i( \hat{\blambda} ) ,
$$
where $\hat{\blambda}'=(\hat{\lambda}_1$, $\hat{\blambda}_2')$ satisfies $\hat{U}_1(\hat{\blambda})=0$ and $\hat{U}_2( \hat{\blambda})=0$, and
\begin{eqnarray*}
\hat{U}_1( \blambda ) &=&\sum_{i \in S} \hat{p}_i ( \blambda ) \pi_i  - W, \\
\hat{U}_2 ( \blambda ) &=&\sum_{i \in S} \hat{p}_i ( \blambda ) \bx_i  - \bar{\mathbf{X}}_N.
\end{eqnarray*}

Now, under some regularity conditions, we can show that $\hat{\blambda}$ converges in probability to $\blambda^*=(\lambda^*_1, {\blambda^*_2}')'$, where $\lambda_1^*=1/W$ and $\blambda_2^*=\mathbf{0}$. 
Since we can express $\hat{\theta}_{\rm EL} = \sum_{i \in S} \hat{p}_i y_i:= \hat{\theta}_{\rm EL} (\hat{\blambda})$ where $\hat{p}_i =\hat{p}_i( \hat{\blambda} )$, we can apply Taylor linearization around $\blambda=\blambda^*$ to get 
\begin{equation}
\hat{\theta}_{\rm EL} = \hat{\theta}_{\rm EL} ( \blambda^* )  - \beta_1 \hat{U}_1(\blambda^*) - \bbeta_2' \hat{U}_2 ( \blambda^*) + o_p (n^{-1/2}) 
\label{a-res1}
\end{equation} 
where $( \beta_1, \bbeta_2')$ satisfies 
\begin{equation}
\begin{pmatrix}
E\{\frac{\partial}{\partial \lambda_1 } \hat{U}_1( \blambda^*)\} & E\{ \frac{\partial}{\partial \blambda_2' } \hat{U}_1( \blambda^*) \} \\
 E\{\frac{\partial}{\partial \lambda_1 } \hat{U}_2( \blambda^*)\} & E\{ \frac{\partial}{\partial \blambda_2' } \hat{U}_2( \blambda^*) \} 
\end{pmatrix}
\begin{pmatrix}
\beta_1 \\
\bbeta_2 
\end{pmatrix}
= \begin{pmatrix}
E\{ \frac{\partial}{ \partial \lambda_1} \hat{\theta}_{\rm EL} ( \blambda^*) \} \\
E\{ \frac{\partial}{ \partial \blambda_2} \hat{\theta}_{\rm EL} ( \blambda^*) \} 
\end{pmatrix}
.
\label{beta}
\end{equation}
Thus, using $W=N^{-1}\sum_{i=1}^N \pi_i=n/N + O_p(n^{-1/2})$  , we can express (\ref{a-res1}) as 
\begin{eqnarray}
\hat{\theta}_{\rm EL} &=& \frac{W}{n} \sum_{i \in S} \frac{y_i}{ \pi_i} + \frac{1}{N} \sum_{i=1}^N \left( \pi_i \beta_1 + \bx_i' \bbeta_2 \right) - \frac{W}{n} \sum_{i \in S} \frac{1}{\pi_i}  \left(  \pi_i \beta_1 + \bx_i' \bbeta_2 \right) + o_p (n^{-1/2}) \notag \\
&=& \frac{1}{N} \sum_{i=1}^N \left\{ \left( \pi_i \beta_1 + \bx_i' \bbeta_2 \right) + \frac{ \delta_i}{ \pi_i} \left( y_i  - \pi_i \beta_1 - \bx_i' \bbeta_2\right) 
\right\} + o_p ( n^{-1/2}) .
\label{a-res2}
\end{eqnarray} 
Now, to estimate $\bbeta$, note that (\ref{beta}) can be written as 
\
\begin{eqnarray*} 
E\left\{ \sum_{i \in S} \pi_i^{-2} \begin{pmatrix} 
\pi_i  - W \\ \bx_i -  \bar{\mathbf{X}}_N 
\end{pmatrix}
 \begin{pmatrix} 
  \pi_i - W  \\ \bx_i -  \bar{\mathbf{X}}_N 
\end{pmatrix}' \right\} \begin{pmatrix} 
\beta_1 \\ \bbeta_2 
\end{pmatrix}
&=& E\left\{ \sum_{i \in S} \pi_i^{-2} \begin{pmatrix} 
\pi_i - W \\ \bx_i -  \bar{\mathbf{X}}_N 
\end{pmatrix}
 y_i  \right\}. \end{eqnarray*}
Therefore, (\ref{result1}) is proved.

\subsection*{B. Proof of (\ref{result2})}  

The consistency of $\hat{\phi}$ to $\phi_0$ can be obtained by showing  $E\{ \hat{U}_b( \phi_0) \}=0$ with some regularity conditions.    
Let $\hat{\theta}_{\rm PS} ( \hat{\phi})=N^{-1} \sum_{i \in S} y_i/ \pi(\bx_i, y_i; \hat{\phi})$.
By applying Taylor linearization,  we obtain 
\begin{eqnarray*}
\hat{\theta}_{\rm PS} &=& \hat{\theta}_{\rm PS} ( \phi_0) +  \left[ E\left\{ \frac{\partial}{ \partial \phi' } \hat{\theta}_{\rm PS} (\phi) \right\} \right] \left( \hat{\phi} - \phi_0 \right) + o_p ( n^{-1/2})\\
&=& 
\frac{1}{N} \sum_{i \in S} \frac{y_i}{\pi_i} -  \left[ E\left\{   Y \mathbf{h}_0 (X,Y)' \right\} \right]\left( \hat{\phi} - \phi_0 \right) + o_p ( n^{-1/2}), \end{eqnarray*}
where $\mathbf{b}_i = \mathbf{b}( \bx_i ; \phi)$ and $
\mathbf{h}_0 (\bx, y; \phi) = \mathbf{h}(\bx,y; \phi) \{ 1-\pi(\bx,y; \phi)
\}$. 

Also,   by Taylor linearization of $\hat{U}_b(\hat{\phi})=0$, we obtain 
 \begin{eqnarray*} 
 \hat{\phi} - \phi_0 &=&  - \left[ E\left\{ \frac{\partial}{ \partial \phi' } \hat{U}_b (\phi_0) \right\} \right]^{-1}\hat{U}_b (\phi_0) + o_p (n^{-1/2}) \\
 &=&  \left[ E\left\{   \mathbf{b} (X) \mathbf{h}_0 (X,Y)' \right\} \right]^{-1} \frac{1}{N}\sum_{i=1}^N \left\{ \frac{\delta_i}{ \pi(\bx_i, y_i ; \phi) } -1 \right\} \mathbf{b}_i   + o_p(n^{-1/2}) .\end{eqnarray*}
 Thus, combining the two results, we can obtain 
 $$
 \hat{\theta}_{\rm PS} = 
\frac{1}{N} \sum_{i \in S} \frac{y_i}{\pi_i} - \frac{1}{N}\sum_{i=1}^N \left\{ \frac{\delta_i}{ \pi(\bx_i, y_i ; \phi) } -1 \right\} \bgamma' \mathbf{b}_i + o_p (n^{-1/2}),$$
where 
$$ \bgamma' =\left[ E\left\{   Y \mathbf{h}_0 (X,Y)' \right\} \right]\left[ E\left\{   \mathbf{b} (X) \mathbf{h}_0 (X,Y)' \right\} \right]^{-1} .$$
Therefore, (\ref{result2}) is established. 
\newpage 

\bibliographystyle{chicago}
\bibliography{ref}

\begin{thebibliography}{}

\bibitem[\protect\citeauthoryear{Baker, Brick, Bates, Battaglia, Couper, Dever,
  Gile, and Tourangeau}{Baker et~al.}{2013}]{baker13}
Baker, R., J.~M. Brick, N.~A. Bates, M.~Battaglia, M.~P. Couper, J.~A. Dever,
  K.~J. Gile, and R.~Tourangeau (2013).
\newblock Summary report of the {AAPOR} task force on non-probability sampling.
\newblock {\em Journal of Survey Statistics and Methodology\/}~{\em 1},
  90--143.

\bibitem[\protect\citeauthoryear{Berger and Torres}{Berger and
  Torres}{2016}]{berger2016}
Berger, Y.~G. and O.~D.~L.~R. Torres (2016).
\newblock An empirical likelihood approach for inference under complex sampling
  design.
\newblock {\em Journal of the Royal Statistical Society, Series~B\/}~{\em 78},
  319--341.

\bibitem[\protect\citeauthoryear{Cao, Tsiatis, and Davidian}{Cao
  et~al.}{2009}]{cao2009improving}
Cao, W., A.~A. Tsiatis, and M.~Davidian (2009).
\newblock Improving efficiency and robustness of the doubly robust estimator
  for a population mean with incomplete data.
\newblock {\em Biometrika\/}~{\em 96}, 723--734.

\bibitem[\protect\citeauthoryear{Chen, Sitter, and Wu}{Chen
  et~al.}{2002}]{chen2002}
Chen, J., R.~Sitter, and C.~Wu (2002).
\newblock Using empirical likelihood method to obtain range restricted weights
  in regression estimators for surveys.
\newblock {\em Biometrika\/}~{\em 89}, 230--237.

\bibitem[\protect\citeauthoryear{Chen and Haziza}{Chen and
  Haziza}{2017}]{chenhaziza2017}
Chen, S. and D.~Haziza (2017).
\newblock Multiply robust imputation procedures for the treatment of item
  nonresponse in surveys.
\newblock {\em Biometrika\/}~{\em 104}, 439--453.

\bibitem[\protect\citeauthoryear{Cochran}{Cochran}{1977}]{cochran77}
Cochran, W.~G. (1977).
\newblock {\em Sampling Techniques\/} (3rd ed.).
\newblock John Wiley {\&} Sons.

\bibitem[\protect\citeauthoryear{Copas and Eguchi}{Copas and
  Eguchi}{2001}]{copas01}
Copas, J.~B. and S.~Eguchi (2001).
\newblock Local sensitivity approximations for selectivity bias.
\newblock {\em Journal of the Royal Statistical Society: Series B\/}~{\em 63},
  871--895.

\bibitem[\protect\citeauthoryear{Copas and Li}{Copas and Li}{1997}]{copas97}
Copas, J.~B. and H.~G. Li (1997).
\newblock Inference for non-random samples.
\newblock {\em Journal of the Royal Statistical Society: Series B\/}~{\em 59},
  55--95.

\bibitem[\protect\citeauthoryear{Dever and Valliant}{Dever and
  Valliant}{2016}]{dever2016}
Dever, J.~A. and R.~Valliant (2016).
\newblock General regression estimation adjusted for undercoverage and
  estimated control totals.
\newblock {\em Journal of Survey Statistics and Methodology\/}~{\em 4},
  289--318.

\bibitem[\protect\citeauthoryear{Elliott and Valliant}{Elliott and
  Valliant}{2017}]{elliott2017inference}
Elliott, M. and R.~Valliant (2017).
\newblock Inference for nonprobability samples.
\newblock {\em Statistical Science\/}~{\em 32\/}(2), 249--264.

\bibitem[\protect\citeauthoryear{Firth and Bennett}{Firth and
  Bennett}{1998}]{firth1998}
Firth, D. and K.~E. Bennett (1998).
\newblock Robust models in probability sampling.
\newblock {\em Journal of the Royal Statistical Society, Series~B\/}~{\em 60},
  3--21.

\bibitem[\protect\citeauthoryear{Franks, Airoldi, and Rubin}{Franks
  et~al.}{2022}]{franks2022}
Franks, A.~M., E.~M. Airoldi, and D.~B. Rubin (2022).
\newblock Nonstandard conditionally specified models for nonignoable missing
  data.
\newblock {\em Proceedings of the National Academy of Science\/}~{\em 117},
  19045--19053.

\bibitem[\protect\citeauthoryear{Fuller}{Fuller}{2009}]{fuller09}
Fuller, W.~A. (2009).
\newblock {\em Sampling Statistics}.
\newblock Hoboken, NJ: John Wiley \& Sons, Inc.

\bibitem[\protect\citeauthoryear{Han and Wang}{Han and Wang}{2013}]{han2013}
Han, P. and L.~Wang (2013).
\newblock Estimation with missing data: Beyond double robustness.
\newblock {\em Biometrika\/}~{\em 100}, 417--430.

\bibitem[\protect\citeauthoryear{Kim}{Kim}{2009}]{kim2009}
Kim, J.~K. (2009).
\newblock Calibration estimation using empirical likelihood in survey sampling.
\newblock {\em Statistica Sinica\/}~{\em 19}, 145--158.

\bibitem[\protect\citeauthoryear{Kim and Shao}{Kim and
  Shao}{2021}]{kimshao2021}
Kim, J.~K. and J.~Shao (2021).
\newblock {\em Statistical Methods for Handling Incomplete Data\/} (2nd ed.).
\newblock CRC press.

\bibitem[\protect\citeauthoryear{Kim and Skinner}{Kim and
  Skinner}{2013}]{kim2013}
Kim, J.~K. and C.~J. Skinner (2013).
\newblock Weighting in survey analysis under informative sampling.
\newblock {\em Biometrika\/}~{\em 100}, 358--398.

\bibitem[\protect\citeauthoryear{Kim and Yu}{Kim and Yu}{2011}]{kimyu2011}
Kim, J.~K. and C.~L. Yu (2011).
\newblock A semi-parametric estimation of mean functionals with non-ignorable
  missing data.
\newblock {\em Journal of the American Statistical Association\/}~{\em 106},
  157--165.

\bibitem[\protect\citeauthoryear{Kott and Chang}{Kott and Chang}{2010}]{kott10}
Kott, P.~S. and T.~Chang (2010).
\newblock Using calibration weighting to adjust for nonignorable unit
  nonresponse.
\newblock {\em Journal of the American Statistical Association\/}~{\em 105},
  1265--1275.

\bibitem[\protect\citeauthoryear{Morikawa and Kim}{Morikawa and
  Kim}{2021}]{morikawa2018}
Morikawa, K. and J.~K. Kim (2021).
\newblock Semiparametric optimal estimation with nonignorable nonresponse data.
\newblock {\em Annals of Statistics\/}~{\em 49}, 2991--3014.

\bibitem[\protect\citeauthoryear{Morikawa, Kim, and Kano}{Morikawa
  et~al.}{2017}]{morikawa2017}
Morikawa, K., J.~K. Kim, and Y.~Kano (2017).
\newblock Semiparametric maximum likelihood estimation under nonignorable
  nonresponse.
\newblock {\em Canadian Journal of Statistics\/}~{\em 45}, 393--409.

\bibitem[\protect\citeauthoryear{Pfeffermann and Sikov}{Pfeffermann and
  Sikov}{2011}]{pfeffermann11}
Pfeffermann, D. and A.~Sikov (2011).
\newblock Imputation and estimation under nonignorable nonresponse in household
  surveys with missing covariate information.
\newblock {\em Journal of Official Statistics\/}~{\em 27}, 181--209.

\bibitem[\protect\citeauthoryear{Pfeffermann and Sverchkov}{Pfeffermann and
  Sverchkov}{1999}]{Pfeffermann1999}
Pfeffermann, D. and M.~Sverchkov (1999).
\newblock Parametric and semiparametric estimation of regression models fitted
  to survey data.
\newblock {\em Sankhy\=a, Series B\/}~{\em 61}, 166--186.

\bibitem[\protect\citeauthoryear{Qin, Leung, and Shao}{Qin
  et~al.}{2002}]{qin2002}
Qin, J., D.~Leung, and J.~Shao (2002).
\newblock Estimation with survey data under non-ignorable nonresponse or
  informative sampling.
\newblock {\em Journal of the American Statistical Association\/}~{\em 97},
  193--200.

\bibitem[\protect\citeauthoryear{Rao}{Rao}{1994}]{rao1994}
Rao, J.~N.~K. (1994).
\newblock Estimating totals and distribution functions using auxiliary
  information at the estimation stage.
\newblock {\em Journal of Official Statistics\/}~{\em 10}, 153--165.

\bibitem[\protect\citeauthoryear{Riddles, Kim, and Im}{Riddles
  et~al.}{2016}]{riddles15}
Riddles, M.~K., J.~K. Kim, and J.~Im (2016).
\newblock Propensity-score-adjustment method for nonignorable nonresponse.
\newblock {\em Journal of Survey Statistics and Methodology\/}~{\em 4},
  215--245.

\bibitem[\protect\citeauthoryear{Rubin}{Rubin}{1976}]{rubin1976}
Rubin, D.~B. (1976).
\newblock Inference and missing data.
\newblock {\em Biometrika\/}~{\em 63\/}(3), 581--592.

\bibitem[\protect\citeauthoryear{{S\"arndal}, Cassel, and Wretman}{{S\"arndal}
  et~al.}{1992}]{sarndal92}
{S\"arndal}, C.~E., C.~M. Cassel, and J.~H. Wretman (1992).
\newblock {\em Model Assisted Survey Sampling}.
\newblock New York: Springer-Verlag.

\bibitem[\protect\citeauthoryear{Shao and Wang}{Shao and Wang}{2016}]{shao2016}
Shao, J. and L.~Wang (2016).
\newblock Semiparametric inverse propensity weighting for nonignorable missing
  data.
\newblock {\em Biometrika\/}~{\em 103}, 175--187.

\bibitem[\protect\citeauthoryear{Till{\'e}}{Till{\'e}}{2020}]{tille2020}
Till{\'e}, Y. (2020).
\newblock {\em Sampling and Estimation from finite populations}.
\newblock John Wiley \& Sons.

\bibitem[\protect\citeauthoryear{Uehara, Lee, and Kim}{Uehara
  et~al.}{2023}]{uehara2020}
Uehara, M., D.~Lee, and J.~K. Kim (2023).
\newblock Semiparametric response model with nonignorable nonresponse.
\newblock {\em Scandinavian Journal of Statistics\/}.
\newblock doi: 10.1111/sjos.12652.

\bibitem[\protect\citeauthoryear{Wang, Shao, and Kim}{Wang
  et~al.}{2014}]{wang14}
Wang, S., J.~Shao, and J.~K. Kim (2014).
\newblock An instrument variable approach for identification and estimation
  with nonignorable nonresponse.
\newblock {\em Statistica Sinica\/}~{\em 24}, 1097--1116.

\bibitem[\protect\citeauthoryear{Wu and Sitter}{Wu and
  Sitter}{2001}]{wu2001model}
Wu, C. and R.~R. Sitter (2001).
\newblock A model-calibration approach to using complete auxiliary information
  from survey data.
\newblock {\em Journal of the American Statistical Association\/}~{\em
  96\/}(453), 185--193.

\bibitem[\protect\citeauthoryear{Yang and Kim}{Yang and Kim}{2020}]{yang2020}
Yang, S. and J.~K. Kim (2020).
\newblock Statistical data integration in survey sampling: A review.
\newblock {\em Japanese Journal of Statistics and Data Science\/}~{\em 3},
  625--650.

\end{thebibliography}

\end{document}